\documentclass{aa}
\usepackage{graphicx}
\begin{document}

\thesaurus{03
              (
               04.19.1;
               11.05.2;
               11.16.1;
               11.19.7;
               12.12.1;
               13.09.1)}

\title{A wide field survey at the Northern Ecliptic Pole I:\\
Number counts and angular correlation functions in K
\thanks{Based on observations collected at the German-Spanish Astronomical
Centre, Calar Alto, operated by the Max-Planck-Institut f\"ur Astronomie,
Heidelberg, jointly with the Spanish National Commission for Astronomy}}
\titlerunning{Number counts and angular correlation functions in K}
\author{M.W.\ K\"ummel\inst{1, }\inst{2}
\and S.J.\ Wagner\inst{1, }\thanks{Email:swagner@lsw.uni-heidelberg.de}}
\authorrunning{M.~K\"ummel \and S.~Wagner}
\institute{Landessternwarte Heidelberg-K\"onigstuhl, K\"onigstuhl 12,
D-69117 Heidelberg, Germany \and (present address:) Max-Planck-Institut f\"ur Astronomie,
K\"onigstuhl 17, D-69117 Heidelberg, Germany}
\offprints{M.W.\ K\"ummel}
\mail{Kuemmel@mpia-hd.mpg.de}
\date{Received 5 May 1999/ Accepted November 5}
\maketitle
\begin{abstract}
We present the results from a multi colour survey performed at the
Northern Ecliptic Pole (NEP). The survey is designed to identify the
counterparts of faint sources from the ROSAT All Sky Survey and the IRAS
survey and to study their optical/near-infrared properties.
We observed the central square degree around the NEP in the optical
bands $B_J$ and $R$ and in the near-infrared band $K$. A shallower
survey was carried out in the optical $I$ band.
Here we present the results of the $K$-band survey.
We discuss the source counts in the magnitude range
$7\,\mathrm{mag}<K<17.5\,\mathrm{mag}$ and the angular correlation
function of galaxies with $K<17.0\,\mathrm{mag}$.
The galaxy counts at the NEP display a subeuklidean slope
in $\mathrm{d}\log N/\mathrm{d}m$. Our shallower slope does
not require the large effects of galaxy evolution or density evolution 
suggested to explain the steeper slopes found in earlier surveys. 
The angular correlation function of galaxies
follows a power law $w^t(\theta) = A\theta^{-\delta}$ with
$A=(5.7\pm0.8)\times 10^{-3}$ and $\delta=0.98\pm0.15$.
This is in accordance with the expected values for stable clustering.
\end{abstract}
\section{Introduction}
Since surveys by satellites obtained deep exposures at the Ecliptic Poles,
the Northern Ecliptic Pole (NEP) is a special location in the sky.
The total exposure time at the NEP in the 
ROSAT All-Sky Survey, for example, was $\le 40\,000\,\mathrm{sec}$ as 
compared to $\le 400\,\mathrm{sec}$ at the ecliptic equator 
(Brinkmann et al.~\cite{brink}).
Likewise, the deepest part of the IRAS survey in the far infrared
($12-100\,\mu\mathrm{m}$) is located at the NEP, reaching flux limits
down to $60\,\mathrm{mJy}$ at $60\,\mu \mathrm{m}$
(Hacking \& Houck \cite{hack}).\\
We took surveys in the optical/near-infrared regime to identify those
sources and to study their broad band energy distribution.
Our surveys cover the central square degree around the NEP and
were carried out in the optical $B_j$- and $R$-band and
in the near-infrared filter $K$. The 95\% completeness limits in $B_j$, $R$, and
$K$ are $24.0\,\mathrm{mag}$, $23.0\,\mathrm{mag}$, and $17.5\,\mathrm{mag}$,
respectively. The results of the identification and details
on our optical surveys will be given in separate papers.
Here we give the number counts of galaxies and point like sources and
present the galaxy two-point correlation function of our $K$-band survey,
which is unprecedented as a homogeneous survey in dynamic range and 
angular size at this depth.\\
Galaxy number counts and galaxy
correlation function are basic quantities in the investigation of
galaxy evolution and galaxy distribution and have both been studied
extensively in the optical regime.
K-band selected samples have the advantage that the K-corrections
are small and nearly independent
of galaxy type (Poggianti \cite{poggi}). Moreover the absolute magnitudes of
galaxies in $K$ are dominated by stars of $M\sim 1\,M_{\sun}$, which makes the
near-infrared as a better tracer of the visible mass of galaxies
as compared to the optical, where the absolute magnitudes may be
strongly affected by star formation.\\
With the completeness limit of $K=17.5\,\mathrm{mag}$ our survey enters a very
interesting magnitude range with a high statistical level given by the large
field. In this magnitude range a break in the slope of the galaxy number
counts from $0.67$ to $0.26$ has been claimed by combining shallow, large
scale surveys with deeper surveys on small fields (Gardner et al. 
\cite{gardner}). From our homogeneous wide, medium-deep survey we can
determine the exact location of this break.\\
Our survey also bridges the gap between correlation functions measured
in deep samples ($K\ge 18.5\,\mathrm{mag}$) which find amplitudes above the
values expected for a non-evolving two-point correlation function
(Roche et al.~\cite{roche98}, \cite{roche99}) and those
values from shallower samples ($K\le 16.0\,\mathrm{mag}$, Baugh et
al.~\cite{baugh}) which are in accordance with stable
clustering (no evolution).\\
\section{Observation and data analysis}
\label{chap:datred}
\subsection{Observation and photometric calibration}
The observations were made with the 2.2-m telescope on Calar Alto, Spain,
during two observing runs from July 22 to 26, 1994 and
August 8 to 10, 1995 using the NIR camera MAGIC (Herbst et al.~\cite{herbst}). 
This camera is equipped with a NICMOS~3 array, similar to the system
described by Hodapp et al.~(\cite{hodapp}) with $256\times 256$ HgCdTe diodes. 
In the {\it Wide Field Mode}, chosen for our project, the field size is 
$6.9'\times 6.9'$ with a scale of $1.6''/\mathrm{pixel}$.\\
The coordinates of the NEP in various coordinate systems are given
in Table~\ref{tab:coo}.
To cover the central square degree around the NEP, we observed a regular
grid of $9\times 9$ fields. On each field several images were taken on 16 
different positions with non-integer pixel offsets in right ascension and 
declination between the individual positions. This dithering allows for 
averaging potential variations of the response on a sub-pixel scale as well 
as for corrections of individual hot pixels, both being essential in
undersampled imaging. In a photometric night we carried out a snapshot
survey where every field was observed in two positions to obtain 
homogeneous photometry. Including the
snapshot survey the total integration time per field is $648\,\mathrm{sec}$.
The data were obtained using a $K'$-filter
(Wainscoat \& Cowie~\cite{wainscoat}) with central wavelength
$\lambda _{central}= 2.10\,\mu \mathrm{m}$ and
$\Delta \lambda = 0.34\,\mu\mathrm{m}$ width (Herbst \cite{herbst95}).\\
In a photometric night we observed NIR standard stars
from the lists of Elias et al.~(\cite{elias}), Zuckermann \& Becklin
(\cite{zucker}), and Casali \& Hawarden (\cite{casali})
and used  the $K$-magnitudes to determine zero point and extinction. Within the
range $-0.036\,\mathrm{mag}\le H-K \le 0.325\,\mathrm{mag}$ covered by the
standard stars, the zero point does not depend on colour in $H-K$.\\
Flux calibration was established using the snapshot survey taken within
three consecutive hours in a single photometric night. These data were
reduced and coadded independently from the main survey to obtain a
homogeneous set of photometric zero-points. This was transferred
to the deeper, coadded survey data for each field individually, using 
several stars in every case.
\begin{table}
\caption{The coordinates of the NEP field centre}
\label{tab:coo}
\begin{tabular}[d]{llll}
\hline
\noalign{\smallskip}
$\alpha_{2000}$:&$18$h$\:00$m$\:00.0$s&$\delta_{2000}$:&$66^{\circ}\:33'\:38.6''$\\
$l_{II}$:&$96^{\circ}.38$&$b_{II}$:&$29^{\circ}.81$ \\
\noalign{\smallskip}
\hline
\end{tabular}
\end{table}
\subsection{Standard reduction}
Standard reduction included dark subtraction, sky subtraction,
and flatfield division of the individual images. Dark exposures were taken
during daytime with the same camera settings that were used for the
scientific exposures, and flatfields were constructed from 
the differences between images of an illuminated and dark screen in the dome.
For every field a sky frame was constructed by scaling the dark corrected
exposures of this field to the same level multiplicatively and removing the 
objects by computing the median for each pixel.\\
The spatial offsets of the images taken on different positions within one 
field were determined using bright stars. Before coadding the images were
resampled to a grid with $0.4''$ pixels.
Due to the resampling it was possible to coadd the images on a
finer grid. This resulted in a better definition of the objects shape and
a stable performance of the source-detection routines.\\
Further analysis was then done only on the central area of each field which
was observed with the full integration time.
This leaves an effective survey area of $0.93\,\mathrm{deg}^2$, spread over
the full square degree with small gaps between adjacent fields.
\begin{table} 
\caption{Completeness and classification parameters}
\begin{tabular}[b]{ccrr}
\hline
\noalign{\smallskip}
Bin & Area [deg$^2$]&$N_{\rm tot}$& $N_{\rm stat}$ \\
\noalign{\smallskip}
\hline
\noalign{\smallskip}
$K<16.0\,\mathrm{mag}$&$0.93\,\mathrm{deg}^2$&3572&4\\
$16.0\,\mathrm{mag}<K<16.5\,\mathrm{mag}$&$0.93\,\mathrm{deg}^2$&1430&10\\
$16.5\,\mathrm{mag}<K<17.0\,\mathrm{mag}$&$0.91\,\mathrm{deg}^2$&2080&120\\
$17.0\,\mathrm{mag}<K<17.5\,\mathrm{mag}$&$0.61\,\mathrm{deg}^2$&1733&371\\
\noalign{\smallskip}
\hline
\end{tabular}
\label{tab:pars}
\end{table}
\subsection{Object detection and photometry}
\label{chap:det}
The detection of sources was carried out with FOCAS (Valdes \cite{valdes},
Jarvis \& Tyson \cite{jarvis}). We tested the reliability of the
detection process as a function of the input parameters on three fields
which were observed with an integration time of $2\,160\,\mathrm{sec}$,
more than three times the amount spent on the other survey fields. 
In these fields we coadded the total of all images to deep frames
as well as subsets of images to frames of integration times
and depths equal to those of the entire survey. Then we performed
source detection on all frames with various sets of detection parameters.
The reliability of the detection was analyzed by identifying the objects
detected on the shallower frames in the deeper ones. 
The detection parameters for the survey were chosen such that only
$<1\,\%$ of the sources in the shallow frames could {\it not} be found in
the deep frames, which gives a reliability of $>99\,\%$ for the sample of
the detected objects. This high reliability was achieved by accepting 
objects above 16 $\sigma$ only.\\
FOCAS performs different types of photometry:
the isophotal flux $L_{i}$ is measured within the detection area,
the so called total flux $L_{total}$, which is determined in the area
expanded by a factor two around the detection area and the flux measured
within a specified aperture around the object, $L_{lfca}$.
$L_{total}$ has the advantage to give good flux estimates independent from
the extent of the objects, because the flux is determined in an area
adapted to object sizes. However, $L_{total}$ misses flux at the faint
level (see e.g.\ Saracco et al.~\cite{saracco}).
To take this into account we chose $L_{total}$ for the bright sources
and $L_{lfca}$, measured in an aperture of $7.2''$ diameter, for the faint
sources. The final value 
depends on the size of the object. If $L_{total}$ is determined in an area
{\it smaller than} $41\,\mathrm{arcsec}^2$ (the area corresponding to the
diameter of the aperture), $L_{lfca}$ is used instead.
The transition from $L_{total}$ to $L_{lfca}$ occurs at
$K \sim 16.5\,\mathrm{mag}$.
We compared the magnitudes of the same objects found on the set of shallow and
deep frames mentioned above to confirm that the
magnitude differences are caused by statistically distributed measurement
errors and did not detect any systematic trend.
\subsection{Morphological classification}
\begin{figure}
\resizebox{\hsize}{!}{\includegraphics*[angle=-90]{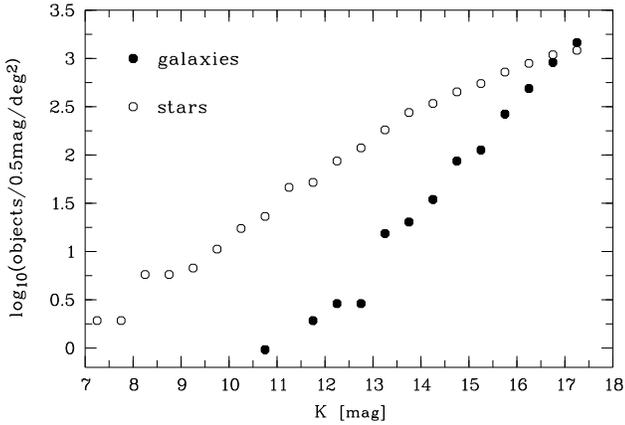}}
\caption{The number density for extended and point-like sources,
labeled as galaxies and stars, respectively.}
\label{Kcoun}
\end{figure}
Since the point spread function (PSF) of the $K$-images is grossly
undersampled, we made no attempt to
classify the objects on the basis of the $K$-data. Instead, the objects
were classified on the basis of the morphological identifying of their 
counterparts in the $R$-survey which has an effective PSF of $\sim 1.2''$.
In the $R$-survey (of which details are given in K\"ummel \cite{drkuemmel} and
K\"ummel \& Wagner in prep.) the classification is based on the
{\it FOCAS resolution classifier} (Valdes \cite{valdes82}).
The classification into point-like and extended objects is reliable for
objects $R\le 21.5\,\mathrm{mag}$. Towards fainter magnitudes
extended objects are progressively misclassified as point-like objects
because their extended parts vanish in the background noise. In the $R$ band
we use a statistical classification by extrapolating the number-counts
of point-like objects ($\mathrm{d}\log N/\mathrm{d}m=const$) into the range of
unreliable classification to estimate the fraction of extended objects.
The (small) fraction of sources detected in $K$ whose $R$-band counterparts 
are too faint for a reliable morphological classification were divided into 
point-like and extended sources based on the fraction of these types at the
corresponding magnitudes of the $R$-band counterparts. This is strictly valid
only if point-like and extended objects have similar R-K colour for
$R \le21.5\,\mathrm{mag}$
\begin{figure}
\resizebox{\hsize}{!}{\includegraphics*[angle=-90]{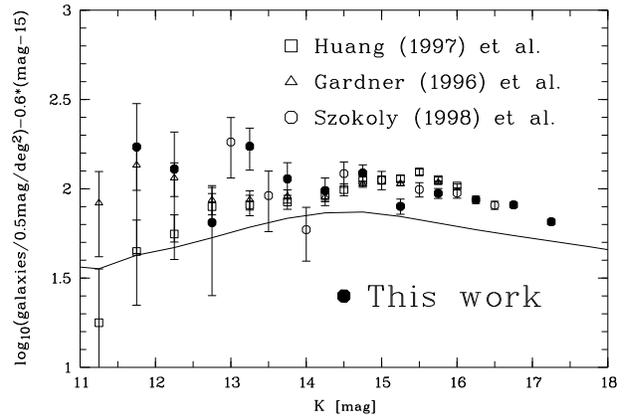}}
\caption{The galaxy counts from the NEP compared to the results
of other surveys covering areas on the square degree scale. The solid line
shows predicted counts from Jimenez \& Kashlinsky (\cite{jimenez})
(their $\Omega_0=0.2$, $\Omega_{\Lambda}=0.0$ model).}
\label{fig:KNcou}
\end{figure}
\subsection{Completeness}
\label{chap:comp}
The completeness was determined using the set of deeper images with 
longer integration times, mentioned in Sect.~\ref{chap:det} which
were used already to find the optimum set of detection
parameters. The object lists of those images were
divided into those found in the shallower images and those
found only in the deep images. This ratio was fitted by a completeness
function $f(mag)$
\begin{equation}
f(mag) = \left[ \exp \left( \frac{mag-mag_{50}}{b}\right) +1 \right]^{-1}
\end{equation}
with $mag_{50}$ being the magnitude where $50\,\%$ of the objects detected in 
the deeper images have been found in the shallower images.
The parameter $b$ is a measure of the width $\Delta mag = mag_2 - mag_1$
where completion drops from
$f(mag_1) = 1-\epsilon$ to $f(mag_2) = \epsilon$. 
We fitted $f(mag)$ to a series of deep/shallow matches taken under
different conditions with different image quality. While $b$ was found to be
independent of image quality ($b=0.26$), $mag_{50}$ varies by
$0.3\,\mathrm{mag}$, reflecting the basic parameters of image quality, i.e.
background noise and FWHM of the PSF.
Establishing the relations between those parameters and $mag_{50}$ and using
these relations and the corresponding zero points 
we computed $mag_{50}$ for every individual survey image. We finally set
a completeness limit $mag_{compl}= mag_{50}-0.5$. Down to this level $f(mag)$ 
was found to be $>0.9$ (i.e. indistinguishable from 1.0) in our deep fields.
We do not use $f(mag)$ to correct counts fainter than $mag_{compl}$ in this
study.
\begin{table}
\caption{Quantitative comparison of galaxy counts}
\label{tab:quantcomp}
\begin{tabular}{lcccc}
\hline
\noalign{\smallskip}
Survey&area&slope&amplitude&range\\
\noalign{\smallskip}
\hline
\noalign{\smallskip}
NEP$_a$ &$0.93$& $0.56 \pm 0.01$ & $97\pm 4$ &10.5-17.0\\ 
NEP$_b$ &$0.93$& $0.41$ & $175$ &16.5-17.5\\
Szokoly&$0.6$&$0.56\pm 0.02$& $98 \pm 5$&12.75-16.75\\ 
Huang&$9.8$ &$0.65\pm 0.01$& $103\pm 2$&11.0-16.0\\
Gardner&$9.8$ &$0.63\pm 0.01$&$103\pm 2$&10.0-16.0\\
\noalign{\smallskip}
\hline
\end{tabular}
\end{table}
\section{Source counts}
\label{chap:ncou}
\subsection{Star and galaxy counts}
Fig.\ \ref{Kcoun} displays the object counts for point-like and extended
source, labeled as stars and galaxies, respectively.
No attempt was made to correct object counts beyond $mag_{compl}$ as
defined in Sect.~\ref{chap:comp}. The counts were derived only from
fields complete to the specific magnitude. The area of the sub-surveys
complete to various depths are given in Table \ref{tab:pars}.
Furthermore, Table \ref{tab:pars} shows the number of objects with
statistical classification in $R$, $N_{\rm stat}$, and the total object number
in that bin, $N_{\rm tot}$.\\
The star counts are in good agreement with the theoretical expected values from
Wainscoat et al.~(\cite{wainstar}),
calculated for $l_{II}=90^{\circ}.0$, $b_{II}=30^{\circ}.0$, which is in
the vicinity of the NEP (see Table \ref{tab:coo}).\\
For $K<17.0\,\mathrm{mag}$ the galaxy counts are well approximated by a
power law with $\mathrm{d}\log N/\mathrm{d}m = 0.56$.
The counts in the faintest bin differ significantly from this slope
and the two faintest bins result in $\mathrm{d}\log N/\mathrm{d}m = 0.41$,
establishing a break or the beginning of a roll-over in the range
$16.5<K<17.0$.
Both the bright and the faint end slope are {\it below} the Euclidean
value of $\mathrm{d}\log N/\mathrm{d}m = 0.6$.
\subsection{Interpretation of the galaxy number counts}
\label{coucomp}
To compare our galaxy counts with the numbers from other
surveys covering areas $>0.6\,\mathrm{deg}^2$, 
Fig.\ \ref{fig:KNcou} gives our data together with the counts
from Gardner et al.~(\cite{gardner96}), Huang et al.~(\cite{huang}) and
Szokoly et al.~(\cite{szokol}). In Fig.\ \ref{fig:KNcou} the Euclidean slope
$0.6$ has been subtracted from the logarithm of the counts
in order to expand the ordinate and to make differences between the counts
clearly visible.\\
We fitted power laws of the form
\begin{equation}
N(mag) = a*10^{b*(mag-15.0)}
\end{equation}
to the counts plotted in Fig.\ \ref{fig:KNcou} and give the
parameters of those fits as well as quality estimators 
in Table \ref{tab:quantcomp}.
The first two columns give the reference and the survey area (in deg$^2$).
Then the slope $b$, the amplitude $a$, and magnitude range of the fits are
listed.  The counts at the NEP are split into two regimes, referenced as
NEP$_a$ and NEP$_b$, because counts at the faint end (NEP$_b$) do not follow
the extrapolation of the power-law of the $K \le 17.0\,\mathrm{mag}$ objects
(NEP$_a$ in Table \ref{tab:quantcomp}).\\
The normalization of the power-laws for the different galaxy counts with
$K \le 16.5\,\mathrm{mag}$, listed in Table \ref{tab:quantcomp}, agree
well among each other within the errors. The slopes of the power laws differ.
While the surveys of Gardner et al. (\cite{gardner96}) and Huang et al.
(\cite{huang}) find values around $b\approx 0.64$, the galaxy counts
of Szokoly et al. (\cite{szokol}) and at the NEP both result in
the subeuklidean value $b=0.56$. The difference between those two
groups can be seen clearly at the counts in Fig.\ \ref{fig:KNcou}.
Power-law fits to the data points from Gardner et al. (\cite{gardner96})
or Huang et al. (\cite{huang}) have a positive slope, which means a
supereuklidean slope for the pure counts, whereas fits to the
data from Szokoly et al. (\cite{szokol}) and the NEP have
negative slopes corresponding to a
subeuklidean slope for the pure counts in $\mathrm{d}\log N/\mathrm{d}m$.\\
Huang et al.~(\cite{huang}) explain the large slope of their counts
with a local under-density of galaxies. Quantitatively they determine
that the normalization of the luminosity function
doubles its value from $z=0$ to $0.2$. Phillips \& Turner (\cite{phillips})
analyze all $K$-band counts to $K=17.0\,\mathrm{mag}$
known to mid-1997 and find two possible explanations for the steep slope:
Either an evolution of the normalization of the galaxy luminosity
function by a factor $1.7-2.4$ to a redshift of $0.1-0.23$ or an unexpectedly
strong low redshift evolution in $K$ which leads to evolutionary and
cosmological corrections as much as $60\,\%$ larger than accepted values.\\
Our shallow, subeuklidean slope of the galaxy counts at the NEP argues against
a local underdensity of galaxies or strong luminosity evolution.
The value $\mathrm{d}\log N/\mathrm{d}m=0.56$ is even smaller than the range
$0.602-0.623$, which were computed in Huang et al.~(\cite{huang}) for different
luminosity functions and cosmological corrections based on {\it constant}
galaxy luminosities.\\
Gardner et al.~(\cite{gardner}) reported a break in the slope of the
galaxy number counts from $\mathrm{d}\log N/\mathrm{d}m=0.67$ to $0.26$.
That break was claimed to occur between $K\approx 15.0\,\mathrm{mag}$
and $K\approx 18.0\,\mathrm{mag}$.
Our data at the NEP show with high statistical significance that this 
break in slope is less extreme, and starts at $K\approx 16.5\,\mathrm{mag}$ 
with a flattening to 0.41 at $K\approx 17.5\,\mathrm{mag}$. A further
flattening to slopes $<0.3$ must occur at magnitudes
$K>17.5-18.5\,\mathrm{mag}$ to match number densities to deeper
source counts from e.g.\ Moustakas et al. (\cite{moust}) and
Djorgovski et al. (\cite{djorg}).\\
Subeuklidean counts in the magnitude range $K=15.0-17.0\,\mathrm{mag}$ 
are supported by the recent compilation of galaxy evolution models from
Jimenez \& Kashlinsky (\cite{jimenez}). In their model the density
of galaxies is assumed to be independent of redshift.
Their predicted $K$-band counts (for $\Omega_0=0.2$ and $\Omega_{\Lambda}=0.0$)
are shown with the solid line in
Fig.\ \ref{fig:KNcou}. Although there is an offset in the normalization of
$\sim 0.15\,\mathrm{dex}$ with respect to {\it all} data displayed in
Fig.\ \ref{fig:KNcou}, the slope of the predicted counts drops
below euklidean values at $K\geq 14.8\,\mathrm{mag}$.
\begin{table*}[tbp]
\caption{The angular correlation function for the various samples}
\begin{tabular}[h]{rr|r|r|r|r||r|r|r}
\hline
\noalign{\smallskip}
\multicolumn{2}{c|} {Sample selection}  & Number of & \multicolumn{3}{c||} 
{$A$ free, $\delta$ fixed} & \multicolumn{3}{c} {$A$ and $\delta$ free} \\

\noalign{\smallskip} \cline{1-2} \cline{4-9} \noalign{\smallskip}

K [mag] & R-K [mag] & galaxies & $A$ [10$^{-3}$] & $\delta$ & $\sigma^2/A$ &
$A$ [10$^{-3}$] & $\delta$ & $\sigma^2/A$ \\

\noalign{\smallskip}
\hline
\noalign{\smallskip}

12.0 - 16.0 & all     &  491 & 9.97 $\pm$ 2.77 & 0.8 & 1.83 & 2.78 $\pm$ 3.42 &
1.14 $\pm$ 0.27 & 2.84 \\
16.0 - 17.0 & all     & 1234 & 5.75 $\pm$ 1.07 & 0.8 & 1.84 & 2.33 $\pm$ 2.13 &
1.02 $\pm$ 0.21 & 2.44 \\

\noalign{\smallskip}
\hline
\noalign{\smallskip}

12.0 - 17.0 & all & 1725 & 5.70 $\pm$ 0.77 & 0.8 & 1.83 & 2.77 $\pm$ 1.78 & 
0.98 $\pm$ 0.15 & 2.30 \\

\noalign{\smallskip}
\hline
\noalign{\smallskip}

12.0 - 17.0 & $>$3.49 &  858 & 9.23 $\pm$ 1.55 & 0.8 & 1.85 & 3.12 $\pm$ 2.47 &
1.08 $\pm$ 0.18 & 2.81\\
12.0 - 17.0 & $<$3.49 &  867 & 7.30 $\pm$ 1.52 & 0.8 & 1.83 & 3.48 $\pm$ 3.50 &
0.99 $\pm$ 0.23 & 2.32\\

\noalign{\smallskip}
\hline
\end{tabular}
\label{tab:crossco}
\end{table*}
\section{Angular correlation function}
\label{chap:angcorr}
\subsection{The calculation of the angular correlation function}
The angular correlation function $w(\theta)$ is defined by
\begin{equation}
 \mathrm{d}P = N^2[1+w(\theta )]\,\mathrm{d}\Omega _1\,\mathrm{d}\Omega _2,
\label{eqn:angdist}
\end{equation}
where $N$ is the mean density of galaxies per steradian and $\mathrm{d}P$ 
the joint probability of finding one galaxy within the solid angle
$\mathrm{d}\Omega _2$ in the projected distance $\theta$ away from a galaxy
within the solid angle $\mathrm{d}\Omega _1$ (Groth \& Peebles \cite{groth}).
$w(\theta)$ is determined by assembling the normalized histograms of
logarithmic distances for the data sample, $\langle DD\rangle$,
a random sample, $\langle RR\rangle$, and for the data-random sample,
$\langle DR\rangle$, and applying (Landy \& Szalay \cite{landy}):
\begin{equation}
w(\theta) = \frac{\displaystyle \langle DD\rangle-2\langle 
DR\rangle+\langle RR\rangle}{\displaystyle \langle RR\rangle}
\label{eqn:firstac}
\end{equation}
The measured function $w^m(\theta )$ computed in Eq.~(\ref{eqn:angdist})
is biased towards low values with respect to the true function,
$w^t(\theta )$, by
\begin{equation}
w^m(\theta) = w^t(\theta) -\sigma^2
\end{equation}
(Baugh et al.\ \cite{baugh}, Roche \& Eales \cite{roche},
Lidman \& Peterson \cite{lidman}),
but this bias $\sigma ^2$ can be estimated as
\begin{equation}
\sigma^2\approx \frac{1}{\Omega^2}\int w^t(\theta_{12}) d\Omega_1\,d\Omega_2
\label{eqn:sigma}
\end{equation}
(Groth \& Peebles \cite{groth}).
The angular correlation function is usually described by a power-law 
$w^t(\theta) = A\theta^{-\delta}$.
If the exponent $\delta$ is known, the amplitude
of $w^t(\theta)$ is determined by fitting the measured correlation to
the function
\begin{equation}
w^m(\theta) = A(\theta^{-\delta}-\sigma^2/A).
\label{eqn:twocorr}
\end{equation}
If the value for $\delta $ is not known, Eq.~(\ref{eqn:sigma}) and
Eq.~(\ref{eqn:twocorr}) have to be iterated.
\subsection{$w(\theta)$ of galaxies $K<17\,\mathrm{mag}$}
\label{chap:angcocol}
\begin{figure}
\resizebox{\hsize}{!}{\includegraphics*[angle=-90]{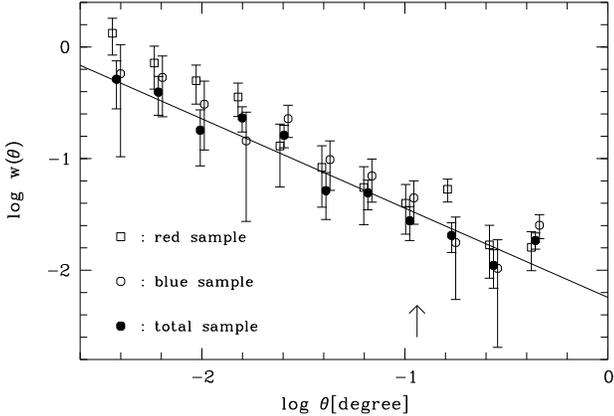}}
\caption{$w^t(\theta)$ for the $K\le 17.0\,\mathrm{mag}$ and for 
blue and the red subsamples. Abscissa values of the
red and the blue sample were offset by $\Delta \log (\theta) = \pm 0.02$.
The solid line displays the power law fitted to the total sample
($\delta =0.8$) and the arrow indicates the separation between two adjacent
fields}
\label{fig:cr}
\end{figure}

We calculated the angular correlation function of galaxies brighter
than $K=17\,\mathrm{mag}$. To avoid spurious detection of the filaments
around the planetary nebulae NGC\,6543 a sufficiently large region
was masked out, leaving a sample of 1725 objects in an area of
$0.9\,\mathrm{deg}^2$.
Following Baugh et al.~(\cite{baugh}) we use
\begin{equation}
\delta w^m(\theta) = 2*\sqrt{(1+w^m(\theta))/[DD])}
\end{equation}
to derive errors, were $[DD]$ denotes the {\it non} normalized histogram
of $\langle DD \rangle$.\\
Table \ref{tab:crossco} give the results
of fitting $w^m(\theta)$ to Eq.~(\ref{eqn:twocorr})
in case for a fixed $\delta=0.8$ and for $\delta$ being a free parameter.
This was computed for the total sample as well as for subsamples
split at $K=16\,\mathrm{mag}$.\\
Fig.\ \ref{fig:cr} shows the bias-corrected angular correlation
function of the total sample with the power law fitted to the data
($\delta =0.8$) on scales $10''<\theta <33'$. While the lower limit is
dictated by the small number of pairs in $[DD]$,
the upper limit is set to half the largest size of the survey field.
The measured correlation follows the power law very well on all scales.
There is no indication of a break or discontinuity on the scale of the distance
between two adjacent fields (marked by the arrow in Fig.\ \ref{fig:cr}).
Even the largest deviation at the widest angular scale ($\theta = 0.44^{\circ}$) 
is at the level of $2.5\,\sigma $, i.e. not significant.\\
With a power-law slope of $\delta \sim 1.0$, and range of more than two orders
of magnitude in $\theta$, a strong variation of the surface density about the
mean value of $0.5\,\mathrm{arcmin}^{-2}$ is implied. Due to the still low
average surface density, the peaks in the surface density distribution
cannot be attributed to clusters in an unambiguous way.\\
We investigated whether the angular correlation depends on galaxy
colour. The total set of objects is divided into a red sample
with $R-K>3.49\,\mathrm{mag}$ and a blue sample with $R-K<3.49\,\mathrm{mag}$.
The two subsamples are selected to include approximately the same number of 
objects (see Table~\ref{tab:crossco}). The angular correlation
function was computed in the same way as for
the total sample and values for the amplitude and scale-length
derived from the fits to the correlation functions 
are given in Table~\ref{tab:crossco}.\\
Fig.\ \ref{fig:cr} shows the bias corrected correlation functions
of the blue and the red sample in comparison to the total sample.
As can be seen in Fig.\ \ref{fig:cr} and in Table~\ref{tab:crossco}
within the errors the blue galaxies show the same clustering properties as
the red galaxies.
\subsection{The angular correlation function of $K$-selected galaxies}
\begin{figure}
\resizebox{\hsize}{!}{\includegraphics*[angle=-90]{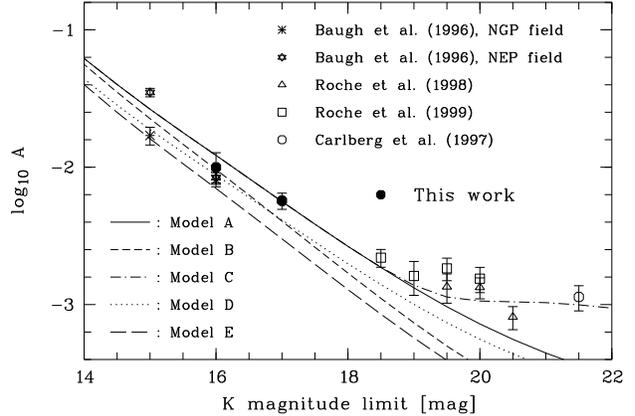}}
\caption{The $w(\theta )$ amplitude of galaxies in our survey at the NEP,
compared with amplitudes from the surveys of Baugh et al.~(\cite{baugh}),
Roche et al.~(\cite{roche98}, \cite{roche99}) and Carlberg
et al.~(\cite{carlberg}). The lines show the theoretical models from
Roche et al.~(\cite{roche98}, \cite{roche99}).}
\label{fig:crmod}
\end{figure}

The two-point correlation function is approximated by a power-law with 
two free variables. It is important to note that these are not derived
independently. The comparison of the amplitude of the correlation is subject
to differences in the angular range involved in the fitting
even for fits with fixed slopes $\delta$.\\
Fig.\ \ref{fig:crmod} gives the change of amplitudes of the angular
correlation function with limiting magnitude in $K$. The values determined in
Sect.\ \ref{chap:angcorr} are displayed together with all other
measurements known to us (from Baugh et al.~\cite{baugh}, Carlberg et
al.~\cite{carlberg}, Roche et al.~\cite{roche98}, \cite{roche99}).
From Baugh et al.~(\cite{baugh}) two values for the amplitude are given in
Fig.\ \ref{fig:crmod} since the amplitude obtained in two different fields
are discrepant at the bright end.\\
The lines in Fig.\ \ref{fig:crmod} show the theoretically expected values
for the amplitude of the angular correlation function as computed
in Roche et al.~(\cite{roche98}, \cite{roche99}) for
various models of the spatial two-point correlation function.
Details of those models can be found in these papers and references therein.\\
The amplitude of our $K<16.0\,\mathrm{mag}$ galaxies is in accordance with
the values from Baugh et al.~(\cite{baugh}) and does not exclude any of the
models A to E due to the large error. The amplitude at the faint level
is in good agreement with the values expected from models A and C.
The difference between the data and models B and D is $2.2\,\sigma$,
thus these models cannot be rejected. Model E is rejected at the
$3.3\,\sigma$ level. We want to point out that the amplitudes compared
in Fig.\ 5 are derived assuming $\delta=0.8$. Piecewise fits of 
$w^m(\theta)$, shown in Fig.\ 3 reveal that narrow angle surveys would
obtain lower normalization and wide angle surveys larger ones. Comparison
of $\mathrm{d}A/\mathrm{d}K$ as shown in Fig.\ 5 rely on $w^m(\theta)$ being well described
by a single power-law.\\
Our data favour model A and C, which were computed assuming 
stable clustering and a significant higher normalization of the
clustering strength for early type galaxies,
as measured by Guzzo et al.~(\cite{guzzo}).\\
A stronger clustering of early type galaxies should result in a
larger angular correlation amplitude for red galaxies. This is in
contradiction to the results from Sect.\ \ref{chap:angcocol}
where the amplitude of the red sample (early type galaxies),
was {\it not} significantly larger than the amplitude of the
blue sample (late type galaxies).
This disagreement could be caused by the diluting effects of the
$K$-band selection of the sample which results in a broader 
distribution of the red galaxies in
$\mathrm{d}N/\mathrm{d}z$ (Roche et al.~\cite{roche99}) and a decrease in
the angular correlation amplitude of the red sample. This would still
imply the existence of several nearby clusters with increased numbers
of early-type galaxies, which are not detected.\\
Finally, we want to stress that further confirmation for the high
correlations (see Fig.\ \ref{fig:crmod}) found in $K>19.0\,\mathrm{mag}$
samples is needed despite existing
measurements of the angular correlation function in samples down to
$K=21.5\,\mathrm{mag}$ (Carlberg et al.~\cite{carlberg}).
Due to the small sample sizes used
in these studies, the derivation of amplitudes (four times smaller
than those found at K$<$17 in the present study) are dominated by
Poisson errors and uncertainties introduced by the correction for 
stellar contamination.
The measurement of the small signals expected at $K\sim 19\,\mathrm{mag}$
would require samples of several $10^4$ sources which are not yet available.
\begin{acknowledgements}
The authors want to thank N.~Roche and R.~Jimenez for providing us in digital
form their model results for the angular correlation function and
galaxy number counts, respectively. This work was supported by the DFG
(Sonderforschungsbereich 328) and the
{\it Studienstiftung des deutschen Volkes}.
\end{acknowledgements}


\begin{thebibliography}{}
\bibitem[1996]{baugh} Baugh C. M., Gardner J. P., Frenk C. S., Sharples R. M.,
1996, MNRAS 283, L15
\bibitem[1999]{brink} Brinkmann W., Chester M., Kollgaard R., et al.,
1999, A\&AS, 134, 221
\bibitem[1997]{carlberg} Carlberg R. G., Cowie L. L., Songaila A., Hu E. M.,
1997, ApJ 484, 538
\bibitem[1995]{djorg} Djorgovski S., Soifer B. T., Pahre M. A., et al.,
1995, ApJ 438, L13
\bibitem[1992]{casali} Casali M., Hawarden T., 1992,
UKIRT Newsletter\ 4, 33 
\bibitem[1982]{elias} Elias J. H., Frogel J. A., Matthews K., Neugebauer G.,
1982, AJ 87, 1029
\bibitem[1993]{gardner} Gardner J. P., Cowie L. L., Wainscoat R. J., 1993,
ApJ 415, L9
\bibitem[1996]{gardner96} Gardner J. P., Sharples R. M., Carrasco B. E.,
Frenk C. S., 1996, MNRAS 282, L1
\bibitem[1977]{groth} Groth E. J., Peebles P. J. E., 1977, ApJ 217, 38
\bibitem[1997]{guzzo} Guzzo L., Strauss M. A., Fisher K. B., Giovanelli R.,
Haynes M., 1997, ApJ 489, 37
\bibitem[1987]{hack} Hacking P., Houck J. R., 1987, ApJS 63, 311
\bibitem[1995]{herbst95} Herbst T. M., 1995, Magic Observer's Guide,
Max-Planck-Institut f\"ur Astronomie, Heidelberg
\bibitem[1993]{herbst} Herbst T. M., Beckwith S. V. W., Birk Ch., et al.,
1993, in: Infrared Detectors and
Instrumentation, ed.~Fowler A. M., SPIE Proceedings Series, Vol.~1946,
p.~605
\bibitem[1992]{hodapp} Hodapp K. W., Rayner J., Irwin E., 1992, PASP 104, 441
\bibitem[1997]{huang} Huang J. S., Cowie L. L., Gardner J. P., et al.,
1997, ApJ 476, 12
\bibitem[1981]{jarvis} Jarvis J. F., Tyson J. A., 1981, AJ 86, 476
\bibitem[1999]{jimenez} Jimenez R., Kashlinsky A., 1999, ApJ 511, 16
\bibitem[1998]{drkuemmel} K\"ummel M. W., 1998, Ph. D. Thesis, University of
Heidelberg
\bibitem[1993]{landy} Landy S. D., Szalay A. S., 1993, ApJ 412, 64
\bibitem[1996]{lidman} Lidman C. E., Peterson B. A., 1996, MNRAS 279, 1357
\bibitem[1997]{moust} Moustakas L. A., Davis M., Graham J. R., et al.,
1997, ApJ 475, 445
\bibitem[1998]{phillips} Phillips L. A., Turner E. L., 1998, submitted to ApJ
(astro-ph/9802352)
\bibitem[1997]{poggi} Poggianti B. M., 1997, A\&AS 122, 399
\bibitem[1998]{roche} Roche N., Eales S. A., 1999, MNRAS 307, 303
\bibitem[1998]{roche98} Roche N., Eales S. A., Hippelein H., 1998,
MNRAS 295, 946
\bibitem[1999]{roche99} Roche N., Eales S. A., Hippelein H.,
Willott C. J.,1999, MNRAS 306, 538
\bibitem[1997]{saracco} Saracco P., Iovino A., Garilli B., Maccagni D.,
Chincarini G., 1997, AJ 114, 887
\bibitem[1998]{szokol} Szokoly G. P., Subbaro M. U., Connolly A. J.,
Mobasher B., 1998, ApJ 492, 452
\bibitem[1982]{valdes82} Valdes F., 1982, in: Instrumentation in
Astronomy IV, ed.\ Crawford D. L., S.P.I.E. Proceedings Series Vol.\ 331,
p. 465
\bibitem[1994]{valdes} Valdes F., 1994, IRAF Group - Central Computer
Services, National Optical Astronomy Observatories, Tucson
\bibitem[1992]{wainscoat} Wainscoat R. J., Cowie L. L., 1992,
AJ 103, 332
\bibitem[1992]{wainstar} Wainscoat R. J., Cohen M., Volk K.,
Walker H. J.,Schwartz D. E., 1992, ApJS 83, 111
\bibitem[1987]{zucker} Zuckerman B., Becklin E. E., 1987, ApJ 319, L99
\end{thebibliography}
\end{document}